\documentclass[aps,pre,amsmath,amssymb,twocolumn]{revtex4}
\usepackage{amsmath,amsthm,amssymb}
\allowdisplaybreaks
\usepackage{graphicx}
\usepackage{epstopdf}
\usepackage{pstricks}
\usepackage{epsf}

\theoremstyle{plain}

\usepackage{graphicx}

\def\pa{\partial}
\newif \ifLastSection \LastSectionfalse

\numberwithin{equation}{section}

\newcommand{\be}{\begin{equation}}
\newcommand{\ee}{\end{equation}}
\newcommand{\ba}{\begin{eqnarray}}
\newcommand{\ea}{\end{eqnarray}}
\newcommand{\baa}{\begin{eqnarray}}
\newcommand{\eaa}{\end{eqnarray}}
\newcommand{\ed}{\end{document}}
\newcommand{\lab}[1]{\label{#1}}
\newcommand{\re}[1]{(\ref{#1})}
\newcommand{\ci}[1]{\cite{#1}}

\begin{document}

\title {Charged solitons  in branched conducting polymers}
\author{D. Babajanov$^{a}$, H. Matyoqubov$^b$,  and D. Matrasulov$^a$}
\affiliation{ $^a$ Turin Polytechnic University in Tashkent, 17
Niyazov Str., 100095,  Tashkent, Uzbekistan\\
$^b$ Urgench State University, 14 H. Olimjon Str., 220100,
Urgench, Uzbekistan}

\begin{abstract}
We consider dynamics of charged  solitons in branched conducting
polymers. An effective model based on the sine-Gordon equation on
metric graphs is used for computing the charge transport and
scattering of charge carriers at the polymer branching points.
Condition for  the ballistic charge carrier transportis revealed.
\end{abstract}
\maketitle

\section{Introduction}
Conducting polymers are the basic materials for organic
electronics, including polymer based photovoltaics. Effective
functionalization of such materials requires understanding  the
mechanisms for charge carrier transport and their optimal control.
Considerable progress has been made in the study of charge
transport mechanisms in conducing polymers during past three
decades \ci{MGH}-\ci{Heeger2}. In particular, several mechanisms
for charge transport, such as exciton, polaron and soliton based
ones have been proposed. Excitons, which are the quasiparticle
consisting of bound electron-hole pair, can be induced e.g., in
conjugated polymers, due to the photon impact. Trapping and
travelling of charge by kink solitons in conducting polymers
provide another mechanism for charge transport. When neutral and
charged solitons appear in bound state, they can form  charge
carriers, which are called polarons. Each mechanism may play key
role depending on the type of a device and functionalization
method. Therefore effective utilization of these mechanisms in
organic electronics requires developing different realistic models
for charge carrier dynamics. Most of the conducting polymers are
typically composed of linear conjugated systems, which can be
observed, e.g., in the  study of polyacetylene, polypyrrole,
polyaniline, polythiophene, and other conjugated polymers.
However, conducting polymers having branched architecture
attracted much attention recently.
\begin{figure}[th!]
\includegraphics[width=80mm]{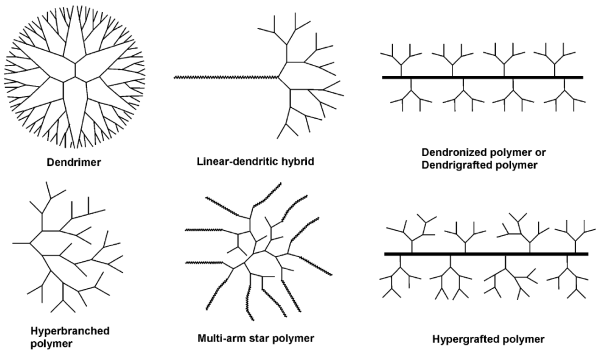}
\caption{Schematic description of dendritic polymers (from the
Ref.\ci{BCP7}).} \label{poly}
\end{figure}
These are kind of polymers in which linear chain  splits into the
two or more branches starting from some point, which is called
branching point, or node, or vertex. The structure of a branching
can have different architecture, e.g. can be in form of star,
tree, ring, etc. These latter implies the rule for branching and
called branching topology. When the topology of a polymer is very
complicated, it is called hyperbranched polymer. Branched polymers
differ from their linear counterparts in several important
aspects. Such polymer forms a more compact coil than a linear
polymer with the same molecular weight. Also, depending on the
topology of branching, electronic and elasticity properties can be
completely different than those of linear polymers. A number of
papers on the synthesis of different branched polymers and study
of their optical, electronic and mechanical properties have been
published in the literature during past two decades (see, e.g.,
\ci{BCP0}- \ci{BCP16}).  A review of the chemistry and physics of
hyperbranched polymers is presented in \ci{BCP3}. In \ci{BCP8} a
synthesis strategy for a hyperbranched sulfonated
polydiphenylamine was developed and electronic properties have
been studied. Synthesis and light-emitting applications of several
conjugated polymers have been reviewed in \ci{BCP9}.  In
\ci{BCP10} the synthesis of dendrimer-like star-branched polymers
is reported. Scalable and versatile synthesis of multifunctional
polyaniline with branched structure with excellent electronic
conductivity and electrochemical properties was reported in
\ci{BCP17}. A review of the synthesis, properties, and
functionality of dendrimers, hyperbranched polymer, and star
polymers with emphasis on functional aspects can be found in
\ci{BCP8}. Synthesis and optical properties of
organosiliconeoligothiophene branched polymers was reported in
\ci{BCP12}. Fabrication and electronic properties of
functionalized branched EDOT-terthiophene copolymer films was
considered in \ci{BCP14}.
\begin{figure}[th!]
\includegraphics[width=80mm]{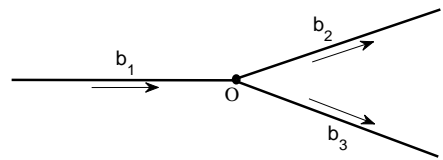}
\caption{Basic star graph.} \label{psg}
\end{figure}

\begin{figure}[th!]
\includegraphics[width=80mm]{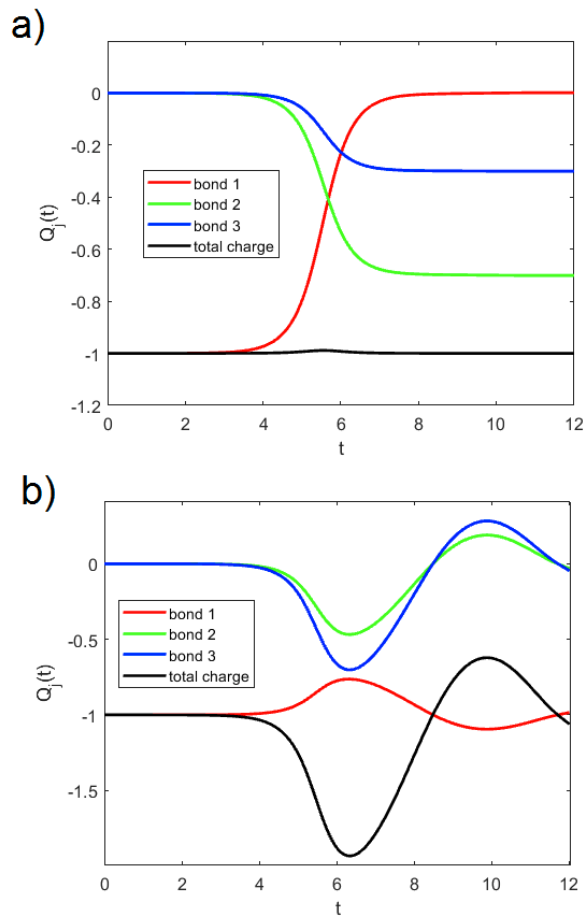}
\caption{Time dependence of charge on each branch of the star
branched polymer: (a) when the sum rule (\ref{r-sum0}) is
fulfilled ($a_1=1,\,a_2=0.7,\,a_3=0.3.$) and (b) broken
($a_1=1,\,a_2=2,\,a_3=3.$) } \label{charge1}
\end{figure}
Some physical properties of   conducting polymer networks are
discussed in \ci{PN1,PN2}. Despite the considerable progress made
in the synthesis and study of branched conducting polymers, the
problem of the charge carrier dynamics in such structures is still
remaining as less studied topic.

In this  paper we consider dynamics of charged solitons  in
branched conducting polymers.  Solitons as charge carriers in
linear conducting polymers have been extensively studied earlier
in the literature \ci{Chsol1} -\ci{Chsol6}. Here we propose a
model based on the sine-Gordon equation on metric graphs to
describe charged soliton transport in branched conducting
polymers. The model considers the branched polymer as a
macroscopic network, where the soliton dynamics is described in
terms of a nonlinear wave equation.
\begin{figure}[th!]
\includegraphics[width=80mm]{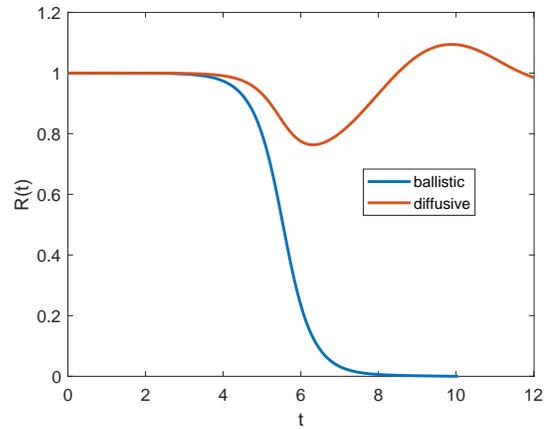}
\caption{Time-dependence of the branching point reflection
coefficient (defined as $R(t)=Q_1(t)/Q_1(t=0)$) for the charged
soliton in star branched polymer in cases of   ballistic and
diffusive transport.} \label{refl}
\end{figure}
Using the exact and numerical solutions of the problem, we study
the problem of charge transport, scattering and transmission of
charged solitons through the branching point. Within the proposed
model, we reveal the conditions for ballistic and diffusive
transport of charged solitons in branched conducting polymers.

We note that  particle \ci{Kost}-\ci{Exner15} and soliton dynamics
in networks has been the topic for extensive research during past
decade (see, e.g., Refs. \ci{Zarif} -\ci{Karim2018}). Modeling
such structures in terms of metric graphs provides powerful tool
for effective description of the wave dynamics in  branched
structures appearing in different areas of physics. The graph
itself is determined as a set of branches which are connected to
each other at the vertices (branching points) according to some
rule. This rule is called the topology of a graph. When branches
of a graph are assigned length it is called matric graph. Topology
of a graph is given in terms of the adjacency matrix which is
defined, e.g., in \cite{Kost,Uzy1,Uzy2,Kuchment04,Exner15}.


Simplest graph topology is called basic star graph, which presents
simple Y-junction with three branches. Advantage of modeling
branched structures by metric graphs is the fact that it allows to
describe the structure as one-dimensional, or  quasi-one
dimensional system. Here we consider star, loop and tree shaped
polymers. This paper is organized as follows. In the next section
we give formulation of the problem together with the description
of the model for star branched polymers. Section III extends to
the model for other branching topologies, such as loop- and
tree-shaped polymers. Finally, section IV presents some concluding
remarks.

\section{Dynamics of sine-Gordon solitons in branched polymers: Modeling for star-shaped topology}

Most of the branched polymers synthesized so far, have star or
dendritic hyperbranched structures, although some other
architectures including those having fractal structure are
available. Some of the hyperbranched  topologies are presented in
Fig. \ref{poly}.
\begin{figure}[th!]
\includegraphics[width=80mm]{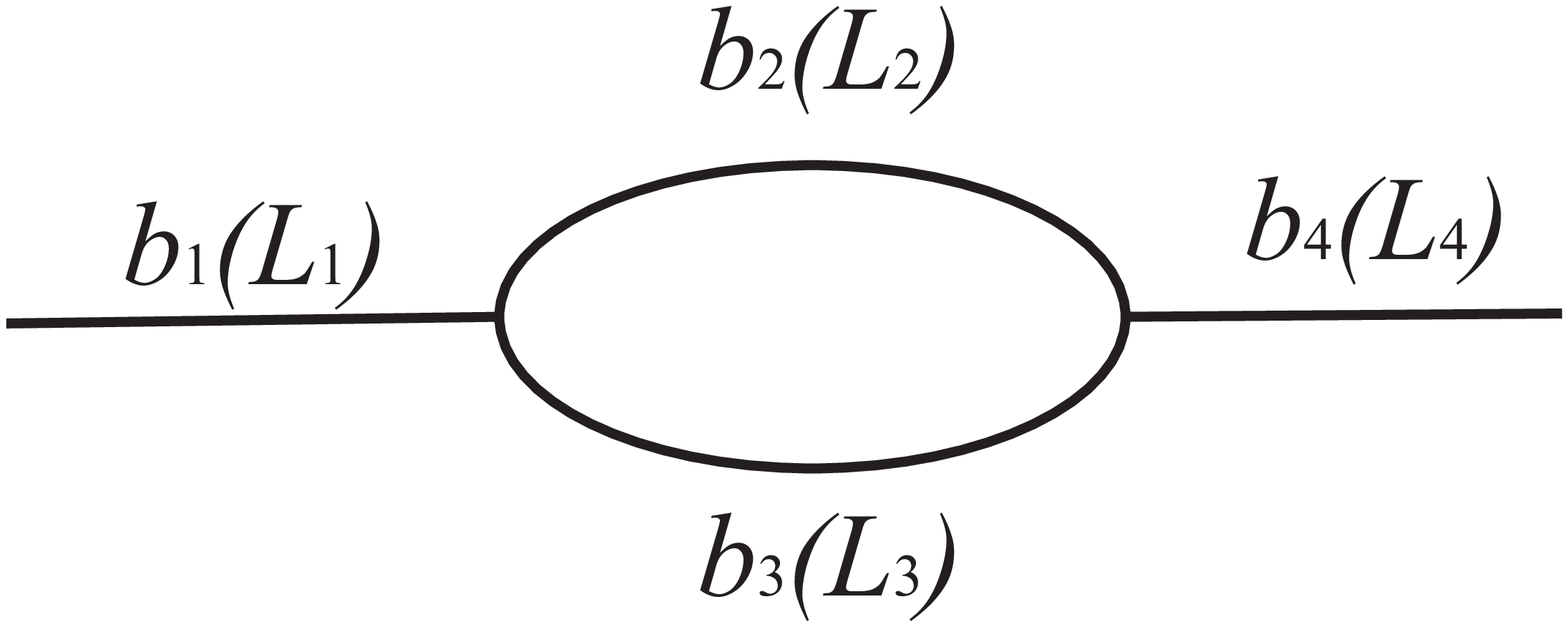}
\includegraphics[width=80mm]{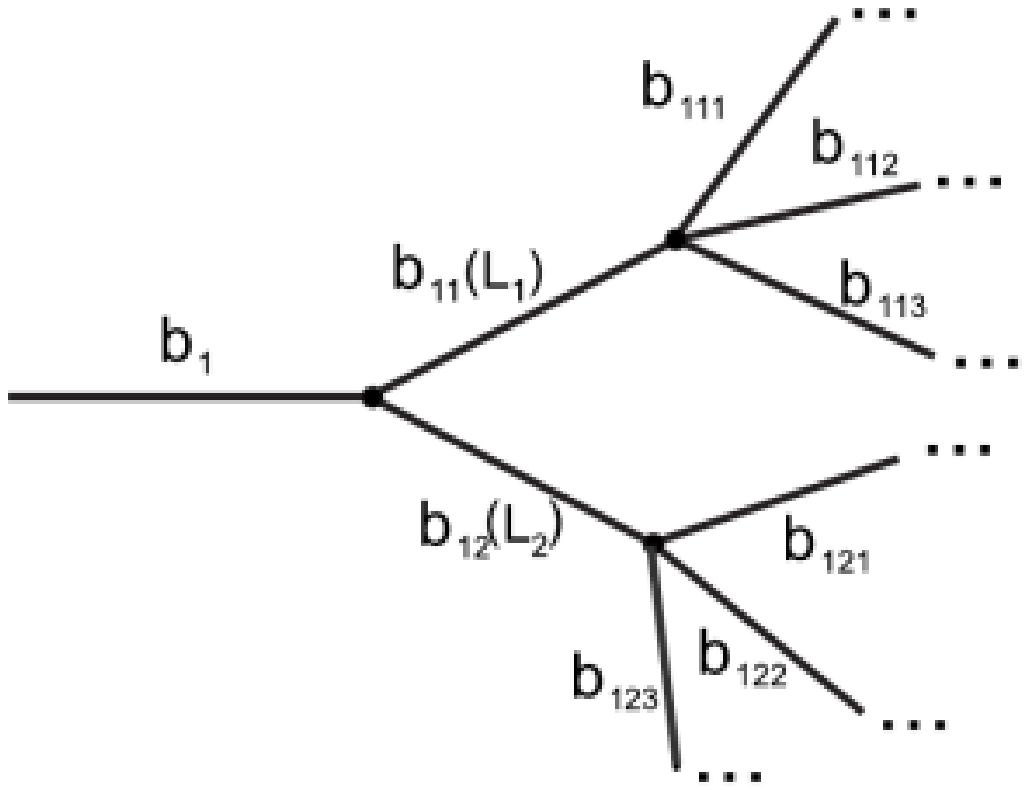}
\caption{Loop (upper) and tree (lower) graphs.} \label{loop}
\end{figure}
There are several models for charged soliton dynamics in
conducting polymers, such as Su-Schrieffer-Heeger \ci{SSH},
Pariser-Parr-Pople \ci{PPP}, time-dependent Hartree-Fock
 \ci{TDHF} models. Here we develop a model that considers charged soliton as
sine-Gordon kink. Within such approach, charged soliton dynamics
in branched polymers can be described in terms of the sine-Gordon
equation on metric graphs. Let us first consider simplest case, a
star branched polymer having the form of Y-junction. We assume
that the branches are very long compared to the thickness of the
polymer chain. Then such polymer can be considered as a basic star
graph with  semi-infinite branches connected at the point $O$
called vertex, or branching point of the graph (see
Fig.\ref{psg}). The coordinates of soliton in such structure  are
defined as $x_1\in(-\infty,0]$ and $x_{2,3}\in[0, \infty)$, where
0 corresponds to the branching point. Dynamics of charged soliton
in such branched polymer can be described in terms of the
sine-Gordon equation  on metric graph which can be written on each
branch as \ci{Our1}
\begin{equation}\label{eq1}
  \psi_{ktt}-a_k^2\psi_{kxx}+\beta_k\sin \psi_k=0.
\end{equation}
To solve this equation, one needs to impose the boundary
conditions at the branching point (vertex) of the graph and
determine the asymptotic of the wave function at the branch ends.
To formulate  vertex boundary conditions (VBC) one can use the
continuity of wave function \ci{Our1}
\begin{equation}\label{r-cont}
  \psi_1(0,t)=\psi_2(0,t)=\psi_3(0,t)
\end{equation}
and fundamental conservation laws such as energy, charge and
momentum conservations. The asymptotic conditions at infinities
can be written as
\begin{figure}[th!]
\includegraphics[width=80mm]{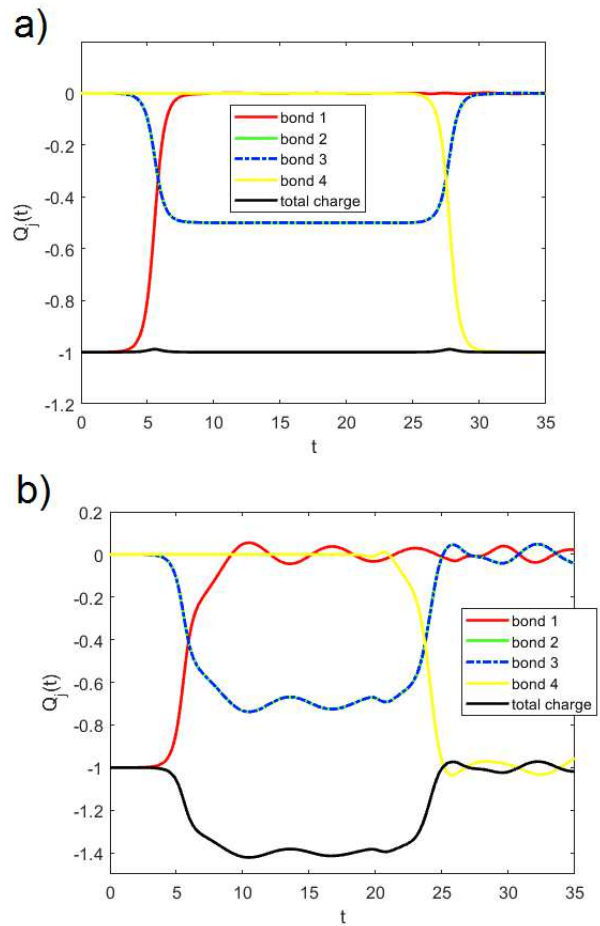}
\caption{Time dependence of charge on each branch of the loop
branched  polymer when (a) the sum rule is fulfilled
($a_1=1,\,a_2=0.5,\,a_3=0.5,\,a_4=1.$) and (b) broken
($a_1=1,\,a_2=0.7,\,a_3=0.7,\,a_4=1).$ } \label{charge1-loop}
\end{figure}
 $\pa_x \psi_1(x_1,t), \pa_t \psi_1(x_1,t)
{\to}0$ and $\psi_1(x_1,t)\to 2\pi n_1$ as $x_1\to-\infty$, and
$\pa_x \psi_k(x_k,t), \pa_t \psi_k(x_k,t) \to 0$ and
$\psi_k(x_k,t)\to 2\pi n_k$ as  $x_{k}\to\infty, k=2,3$, for some
integer $n_k$, $k=1,2,3$.

For the  star graph in Fig \ref{psg}, the energy and topological
charge are defined as (respectively) \ci{Our1}
\begin{equation}\label{energy}
E(t)=\sum_{k=1}^{3}\frac{1}{\beta_k}\int\limits_{B_k}\left[
\frac{1}{2}\left(\psi_{kt}^2+a_k^2\psi_{kx}^2\right)+\beta_k(1-\cos
\psi_k)\right]dx,
\end{equation}
and
\begin{equation}\label{topol-index}
  2\pi
  Q=\frac{a_1}{\sqrt{\beta_1}}\int\limits_{-\infty}^{0}\psi_{1x}dx+\sum_{k=2}^{3}\frac{a_k}{\sqrt{\beta_k}}\int\limits_{0}^{+\infty}\psi_{kx}dx,
\end{equation}
 where  $B_1{=}(-\infty,0), \ B_{2,3}{=}(0,+\infty)$.
From the conservation laws given by
$$
\frac{dE}{dt} = 0, \;\;\; \frac{dQ}{dt} = 0,
$$
\begin{figure}[th!]
\includegraphics[width=80mm]{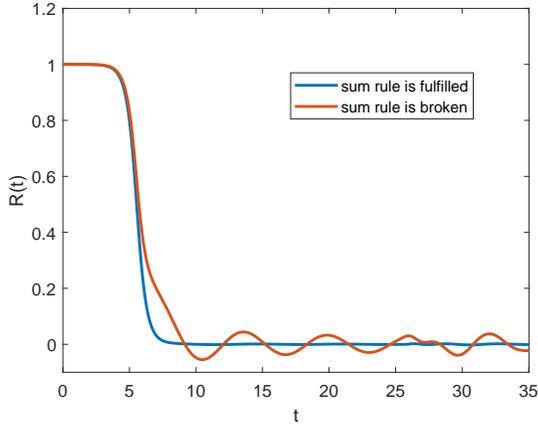}
\caption{Time-dependence of the branching point reflection
coefficient (defined as $R(t)=Q_1(t)/Q_1(t=0)$) for loop branched
polymer in ballistic and diffusive regimes.} \label{refl-loop}
\end{figure}
we have the boundary conditions at the branching point:
\begin{equation}\label{VC-E}
\left.\frac{a_1^2}{\beta_1}\psi_{1x}\right|_{x_1=0}=\left.\frac{a_2^2}{\beta_2}\psi_{2x}\right|_{x_2=0}+
\left.\frac{a_3^2}{\beta_3}\psi_{3x}\right|_{x_3=0}.
\end{equation}
It was shown in \ci{Our1} that exact soliton (kink)solutions of
Eq.\re{eq1} fulfilling the vertex boundary conditions given by
Eqs.\re{r-cont} and \re{VC-E} can be obtained, provided the
following  sum rule holds true:
\begin{equation}\label{r-sum0}
\frac{a_1}{\sqrt{\beta_1}}=\frac{a_2}{\sqrt{\beta_2}}+\frac{a_3}{\sqrt{\beta_3}}.
\end{equation}
Then the solution can be written as
 \be
\psi(x,t) = v(\frac{\sqrt{\beta_k}}{a_k}x,\sqrt{\beta_k}t),
 \ee
 where
\be v(x,t)= 4 \arctan\left[ \exp\left(\pm\frac{x-x_0-\nu
      t}{\sqrt{1-\nu^2}}\right)\right], \lab{kink}
\ee with $|\nu|<1$ being the velocity of the kink.


Thus dynamics of the charged solitons in branched conducting
polymers is described in terms of the problem given by
Eqs.(\ref{eq1}), (\ref{r-cont}) and (\ref{VC-E}).  In other words,
within our model, solution of sine-Gordon equation on such graph
describes the motion of charged kink solitons in branched
conducting polymers. Detailed mathematical treatment of this
problem was given recently in the Ref.\ci{Our1}. Here we apply
these results to the problem of charged soliton transport in
branched conducting polymers by considering different branching
topologies. Having the solution of the problem given by Eqs.
(\ref{eq1}), (\ref{r-cont}) and (\ref{VC-E}), one can compute
different characteristics of the charge transport, such as
time-evolution of the topological charge, reflection and
transmission of charge carrying solitons at the polymer branching
point. In the following for numerical computations we assume that
$\beta_1 =\beta_2 =\beta_2 =1$.
\begin{figure}[th!]
\includegraphics[width=80mm]{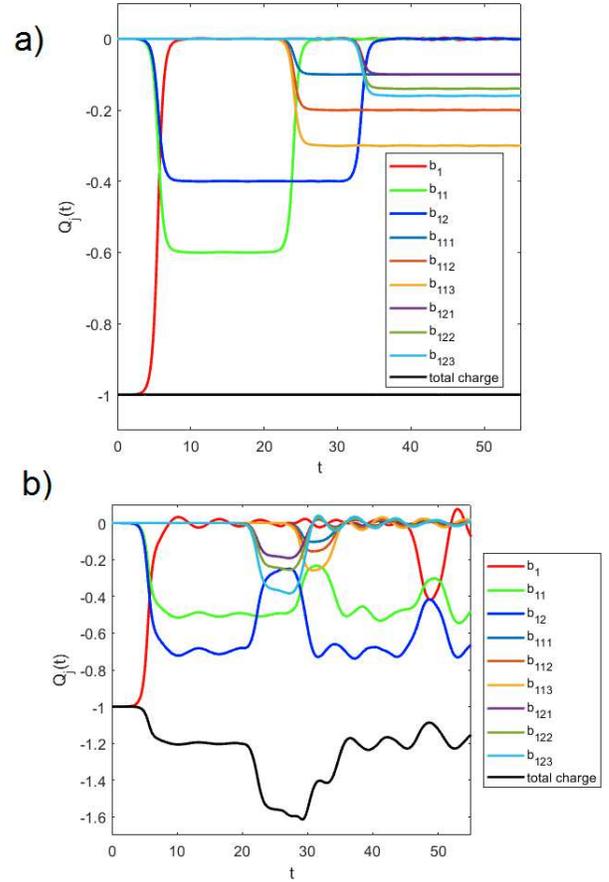}
\caption{Time dependence of charge on each branch of the tree
branched polymer when (a) the sum rule is fulfilled ($a_1=1,
a_{11}=0.6, a_{12}=0.4, a_{111}=0.1, a_{112}=0.2, a_{113}=0.3,
a_{121}=0.1, a_{122}=0.14, a_{123}=0.16$) and (b) broken ($a_1=1,
a_{11}=0.5, a_{12}=0.7, a_{111}=0.2, a_{112}=0.3, a_{113}=0.5,
a_{121}=0.3, a_{122}=0.4, a_{123}=0.6$).} \label{tree1}
\end{figure}
Fig. \ref{charge1}a presents plots of the charge as a function of
time on each branch of the star-branched-polymer for the regime,
when the sum rule given by Eq.\re{r-sum0} is fulfilled, i.e. the
case when energy and charge conservations hold true. Conservation
of topological charge can be clearly seen from this plot. Fig.
\ref{charge1}b presents similar plots for the case, when sum rule
is broken. It is clear that the charge conservation is not
fulfilled in this regime. In both cases it is assumed that charge
is generated on the first branch at the initial time ($t=0$).  In
Fig. \ref{refl} reflection coefficient of charged solitons from
the (star-shaped) polymer branching points are plotted for the
regimes, when the sum rule given by Eq.\re{r-sum0} is fulfilled
and broken. As it can be seen from these plots, there is no
reflection of solitons at the branching point, when the sum rule
is fulfilled. However, for the case, when the sum rule is broken,
one can observe reflection and backward motion of the charge
solitons at the branching point. Such an effect can be effectively
used for tuning of charge transport in branched conducting
polymers and organic electronic devices fabricated on their basis.

\section{Other branching topologies}
The above approach   can be extended for modeling of charged
soliton dynamics in other branched polymers having more
complicated branching topologies. Consider first the loop branched
polymer which consists of two semi-infinite branches connected by
two or more branches having finite lengths $L_1$ and $L-2$. Such
polymer can be considered as a loop graph presented in
Fig.\re{loop}). Under the constraints given as \ci{Our1}
\begin{equation}\label{r4}
a_1=a_1 +a_2=a_{4}
\end{equation}
for the coefficients, and $L_k=a_k L$ ($k=1,2,3,4$)  with a
constant $L$, one can write exact soliton solutions of sine-Gordon
equation on such structure. In other words, the kink solution
given by Eq.\re{kink} is the solution of Eq.\re{eq1} provided
coefficients $a_j$ fulfill the sum rule \re{r4} \ci{Our1}. When
the sum rule is broken, one can solve the problem only
numerically. Having found such solutions, one can compute
different characteristics of the charge transport.
Fig.\re{charge1-loop}a presents plots of time-dependence of the
charge on each branch of the loop branched polymer for the case
when sum rule in Eq.\re{r4} is fulfilled. In
Fig.\re{charge1-loop}b similar plots are presented  for the case,
when sum rule in Eq.\re{r4} is broken. For this case we solve
sine-Gordon equation numerically. In Fig. \ref{ref2} reflection
coefficient of charged solitons from the polymer branching points
are plotted for the regimes, when the sum rule given by
Eq.\re{r-sum0} is fulfilled and broken. Again, the plot confirms
absence of the reflection at the branching point in the regime,
when the sum rule is fulfilled.

Another topology which can be exactly treated within the above
approach is called tree-branching, which corresponds to
hyperbranched conducting polymers.  Such structure can be modeled
in terms of metric tree graph presented in lower panel of Fig.
\ref{loop}. In simplest case, the graph consists of three
``layers'' $b_1, (b_{1i}), (b_{1ij})$, where $i,j$ run over the
given branches. On each branch $b_1, b_{1i}, b_{1ij}$ we have a
sine-Gordon equation given by (\ref{eq1}). Writing on each branch
of the graph Eq.\re{eq1}, setting
$\beta_1=\beta_{1i}=\beta_{1ij}=1$ for all $i,j$, and assuming
that the $a_{1i}$ and $a_{1ij}$  fulfill constraints in the form
of sum rule, explicitly given, e.g. in \ci{Our1}, one can obtain
exact soliton solutions of sine-Gordon equation on the tree graph.
Fig.\ref{tree1}a presents plots of the time-dependence of the
charge on each branch of the tree-shaped conducting polymer,
obtained using the exact solution of sine-Gordon equation on
metric tree graph, i.e., for the case when sum rule for $b_1,$
$b_{1i}$ and $b_{1ij}$ is fulfilled. Similar plots for the case
when the sum rule is broken, is presented in Fig.\ref{tree1}b.
This latter clearly shows breaking of charge conservation law. We
note that according to the conclusion made in \ci{Our1} the above
approach can be used for solving of sine-Gordon equation on metric
graphs with arbitrary topology. The only requirement for the
structure of such polymers is that it contains at least three very
long, outgoing branches.

\section{Conclusions}

We studied charged soliton dynamics in branched conducting
polymers. A model based on the  sine-Gordon equation on metric
graphs is proposed for the analysis of charged  solitons transport
in conducting polymers with different branching topologies.
Time-dependence of the topological charge and kink scattering at
the branching point are analyzed. The regime when there is no
backstattering of charged solitons at the polymer branching points
is revealed. Such regime corresponds to the case, when the
sine-Gordon equation on metric graph is completely integrable and
charge and energy conservation rules are fulfilled. For the
polymers having many branching points, such regime implies
ballistic charge transport.  When the conservation rules are
broken, transmission of charge carriers through the polymer
branched point is accompanied by reflection. Such regime
corresponds to diffusive transport of charged solitons through the
polymer network. Existing the regime, when the transmission of
charge carriers through the branching point is reflectionless
makes possible to achieve very high conductivity  in the polymer
based organic electronic devices. The above model is applicable
for arbitrary branching topology, provided the polymer consists of
at least three very long branches and other branching part is
located between these three branches. Finally, we note that our
model allows to consider also breathers in branched polymers,
using the breather solutions of sine-Gordon equation on metric
graphs.

\end{document}

\begin{equation}\label{r4}
a_1=a\sum_{k=1}^{n}a_k=a_{n+1}
\end{equation}